\documentclass{ws-rv9x6}
\usepackage[square]{ws-rv-van}             

\makeindex
\begin{document}

\chapter[TO UNDERSTAND NATURE - COMPUTER MODELLING ...]{TO UNDERSTAND NATURE - COMPUTER MODELLING BETWEEN GENETICS AND EVOLUTION.}

\author[D. Mackiewicz, S. Cebrat]{Dorota Mackiewicz${}^1$, Stanis{\l}aw Cebrat${}^1$}

\address{${}^1$ Department of Genomics, Faculty of Biotechnology, University of Wroc{\l}aw, ul. Przybyszewskiego 63/77, 51-148 Wroc{\l}aw, Poland. 
E-mail: cebrat@smorfland.uni.wroc.pl.}

\begin{abstract}
We have presented the basic knowledge on the structure of molecules coding
the genetic information, mechanisms of transfer of this information from DNA
to proteins and phenomena connected with replication of DNA. In particular,
we have described the differences of mutational pressure connected with
replication of the leading and lagging DNA strands. We have shown how the
asymmetric replication of DNA affects the structure of genomes, positions of
genes, their function and amino acid composition. Results of Monte Carlo
simulations of evolution of protein coding sequences have shown a specific
role of genetic code in minimizing the effect of nucleotide substitutions on
the amino acid composition of proteins. The results of simulations were
compared with the results of analyses of genomic and proteomic data bases.
This chapter is considered as an introduction to further chapters where
chromosomes with genes represented by nucleotide sequences were replaced by
bitstrings with single bits representing genes.
\end{abstract}

\body

\section{Introduction}
\begin{quotation}
\noindent One day, Leo Szilard informed his friend that he is going to write a diary. 
,,I am not going to publish it but just to inform God about some facts''. 
,,Do you think that God doesn't know facts?'' asked his friend. 
,,I am sure, God doesn't know this version of facts'' answered Leo.	
\end{quotation}
This chapter is a kind of the diary. It is written in a few months but authors try to show their way of understanding the Nature or rather of understanding some aspects of the biological evolution. ,,Their way'' means both - how they understand and what was the chronology of understanding the facts. 
In the first part, the simplest rules of coding the genetic information are described. Everybody knows what is the genetic code\index{genetic code}, everybody knows what is the structure of the DNA double helix, maybe even how it is translated for amino acid sequences in proteins, but it is not a common knowledge how Nature exploits the fact that the two strands of DNA differ in the mechanism of their synthesis, composition, mutation pressure and selection. It is the only part where genes are represented by long series of different elements which have to be translated into chains of other compounds and their biological functions are checked at the end of the process of the virtual evolution. 
In the next parts these genes with very complicated structures are replaced by single bits. Reader can intuitively feel that it is a swindle. Nevertheless, remember that it is not a physicist who tries to convince you that such simplification is possible - it is a biologist. Physicists are very skilful in such simplifications, they were even able to count the planets paths assuming that they are just points - everybody knows that Earth is not a point! Using our simplification that gene is represented by one bit we were able to answer such questions like why men are altruistic and women are egoistic, why women live longer than men, why Y chromosome is short, why Nature invented death and if there is any significant difference between the fruit flies and humans. All answers have been obtained assuming that at least a part of genomic information is switched on chronologically.
In the third part we have even resigned from this chronology, instead, we have introduced some specific behaviour of our virtual creatures, particularly their sexual behaviour. Using a very simplified mechanism mimicking meiosis\index{meiosis} (unfortunately necessary in sexual reproduction as well as in the modelling such reproduction) we were able to show that speciation is a very simple phenomenon even in sympatry. The last results would be interesting even for Darwin himself when writing his ,,The Origin of Species by means of Natural Selection''. 
Since it is a biological paper, rather, and the mathematical formal language is too difficult for biologists, you should not expect formulas in the text. But the text is dedicated for mathematicians and they simply can translate the results obtained by the Monte Carlo methods for their beautiful phenomenological language. I hope that everybody, after reading the text will have much more questions and problems which could be quantitatively described, and readers should judge if God should be informed about such quantitative versions of the biological facts.

\section{Evolution of the DNA coding sequences}
\subsection{DNA double helix}
Before Watson and Crick discovered the DNA (DeoxyriboNucleic Acid) structure, Griffith  had found that there was some information in the bacterial cell which could be transferred from the dead cell to the living cell, changing the (genetic) property of the last one \cite{Grif}. In 1944 Avery's group found that it was probably the DNA molecule responsible for transmitting the information between these cells \cite{Avery}. Avery only warily suggested that it could be the DNA but in fact he proved that. Scientific world was not ready to accept the hypothesis that such a dull molecule, composed of only four different subunits can be a carrier of genetic information. Proteins seemed to be much better and obvious candidates for such an important role. There were other very fundamental discoveries which enabled the discovery of the double helix - the Chargaff's rules\index{Chargaff's rules}. Chargaff found that the number of adenine (A) equals the number of thymine (T) while the number of guanine (G) equals the number of cytosine (C) in any DNA preparation. Moreover, Chargaff found that the ratio $[A+T]/[G+C]$ is constant for a species independently of the tissue where the DNA was isolated from, though the ratio could be different for different species \cite{chargaff}. The letters A, T, G and C will be used in the text for connotation the four nucleotides building the DNA.
Now, more than 50 years after the DNA double helix discovery, it is very easy to say what the rationale should be at the bases for creating the model - in fact, only four of them (Fig. \ref{fig1}):

\begin{figure}
\centering
\includegraphics[width=0.7\textwidth]{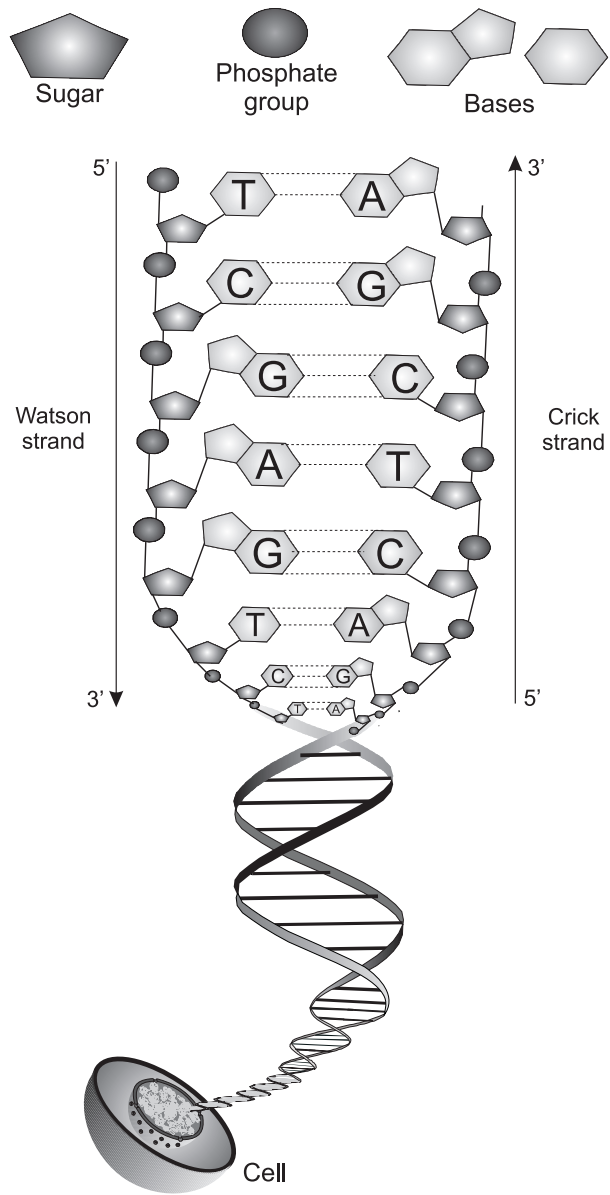}
\caption{\label{fig1} Schematic structure of DNA molecule. Bases: A, T, G and C stand for Adenine, Thymine, Guanine and Cytosine respectively, they form the rungs of the ,,ladder'' while phosphate groups and sugar molecules (deoxyribose) built the rails. Notice two antiparallel strands forming the double helix. The diameter of helix is about 2 nm and length could be measured even in centimetres or several millions of base pairs. }
\end{figure}

\begin{enumerate}
\item the molecule was a double helix;
\item the phosphate backbone was on the outside, bases on the  inside;
\item the strands were antiparallel;
\item specific base pairing keeps the two strands together.
\end{enumerate}

April 25th, 1953 Watson and Crick published in ,,Nature'' a paper entitled ,,A Structure for Deoxyribose Nucleic Acid''. It begins with the sentence ,,We wish to suggest a structure for the salt of deoxyribose nucleic acid (DNA). This structure has novel features which are of considerable biological interest'' \cite{Watson}.
They proposed that the DNA molecule is a double helix that resembles a gently twisted ladder (ten rungs per one twist). The rails of the ladder are made of alternating units of phosphate and the pentose (deoxyribose); the rungs are composed of pairs of nitrogen bases: AT and GC. Here we find this Chargaff's rules, called now the deterministic complementarity rules or parity rules I\index{parity rule} (see Fig. \ref{fig1}). 
The DNA model is very elegant but there is one very peculiar feature of the DNA molecule - resulting from the third assumption - the rails are antiparallel. To be antiparallel means that the rails have their directions. Deoxyribose is an asymmetric molecule with five atoms of carbon (numbered 1' - 5'). Phosphate links the atom number 3' of one deoxyribose with the atom 5' of the other one. It is obvious that at one end there is a free carbon number 5' while at the other end there is a free carbon number 3'. For chemists it is enough to say that the whole DNA strand has a direction 5' to 3'. At the end of the double helix DNA molecule one strand has 3' end and the other one has 5' end. Do you think it is not so important? Imagine, you are in the middle of the molecule and you try to reach the end of it - where is the end? Even if you know that the end is marked by the free 3' carbon - but, of which strand? Thus, it is very important to determine the direction in the double strand or, if we accept the direction of single strand $5'\Rightarrow3'$ then we have to indicate the strand we are thinking about. Please keep in mind this feature because a lot of phenomena discussed in this chapter will show the consequences of this very nuance in the DNA structure for its coding capacity, control of information expression and evolution. 

\subsection{Chromosomes}\index{chromosome}
We will see in the next sections how the information imprisoned in the DNA molecule can be expressed and used by the ,,body'' of the host organism. Now, the only issue which should be important for us is to imagine the DNA molecule inside the compartments where we can find it. If we are talking about the living organisms and we assume that viruses don't live (I am not sure if I know why), then the bacteria are the smallest living creatures which use DNA as their genetic data base. The whole information of one bacterial cell is called its genome. Usually it is built of one DNA molecule though, some species of bacteria posses the genome composed of more than one molecule. If the molecule is indispensable for life of the bacteria cell it is called chromosome, otherwise it is called plasmid. We will talk about chromosomes, only. The free living bacteria usually have larger genomes - up to 10 millions of nucleotides (!).  Some bacteria have found a very comfortable niche for living - the interior of other cells. Those bacteria usually reduce their genome just because they don't need the whole information indispensable for fighting for everything; they have almost everything ready in the host cell. For such bacteria 0.5 million of nucleotides in their data base could be enough. 
Escherichia coli, bacterium living in our alimentary tract is in the middle of this range and has DNA molecule built of about 5 millions of nucleotides. The diameter of the DNA molecule is about 2 nm and the length ... about million times larger - of the order of millimeters. It is about thousand times more than the diameter of the cell. This range of proportion is very common for bacterial as well as eukaryotic cells,  DNA molecule is three orders longer than the diameter of the compartment in which it exist.

\subsection{Genomes}\index{genome}
Here we present some information on the bacterial genomes. Escherichia coli chromosome is circular, like most of bacterial chromosomes. All information necessary for life is enclosed in the chromosome. Nevertheless, bacteria can posses a lot of additional information indispensable for surviving in the special conditions like antibiotics in the environment. This information can be encoded in plasmids, which are usually also circular and sometimes as large as chromosome. The difference between plasmid and chromosome is in the control of their replication.\index{replication} Replication of chromosomes is both, positively and negatively controlled and perfectly synchronized with the cell division and simply it is impossible to imagine the living cell without chromosome - without plasmid it is possible. Thus, we have to replicate the chromosome before we try to divide the cell.

\subsection{Topology of DNA replication}\index{replication}
The topology of DNA replication, even of such a small molecule as E. coli chromosome, is a good opportunity for mathematicians to exercise their imagination - usually the abstract thinking satisfies them entirely. 
Imagine a rope in your living room, 1 cm in diameter and 10 km long, ends tied. The rope is made of two lines wrapped around themselves 500,000 times. During the replication you have to separate the two lines and to connect the new lines with each of the separated one. E. coli can do all that in 45 minutes. During this time it is able to build into the DNA 10 millions new nucleotides (five millions into each of the new strands). To make the exercise more demanding - E. coli can divide every 20 minutes in the favorable conditions and need 45 minutes to replicate its chromosome. After a few generations its genetic information would be diluted. Do you imagine why it is not a case? Because the replication cycles start at 20 minute intervals, chromosome can replicate in many points simultaneously.

\subsection{DNA asymmetry}\index{DNA asymmetry}
\begin{figure}[t!hb]
\centering
\includegraphics[width=0.9\textwidth]{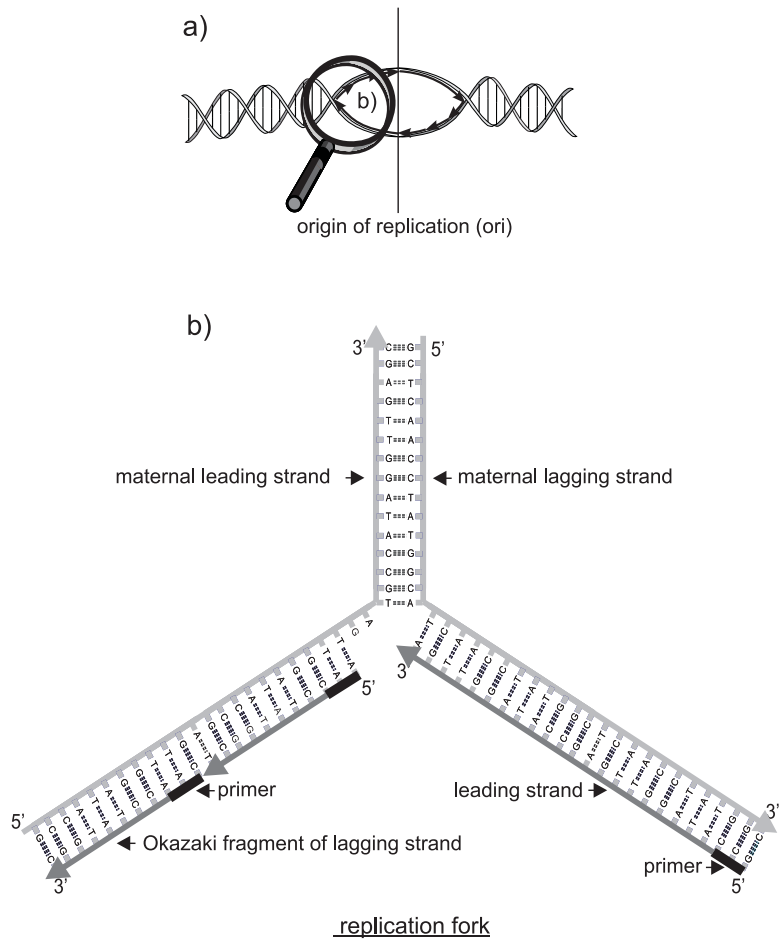}
\caption{\label{fig2} Schematic representation of DNA replication. Panel ,,a'' shows the loop close to origin of replication - oriC where the replication starts in two directions. At this point the two DNA strands have their ,,switches'' from leading to lagging mode of replication. The lower panel shows the differences in mechanisms and topology of replication of the leading and the lagging DNA strands in the region of replication fork. The two replication forks shown in the upper panel will meet at the opposite pole of circular DNA chromosome - terminus of replication. }
\end{figure}
The second problem is connected directly with the mechanism of DNA replication. Replication is performed by the specific biochemical machinery with a speed higher than 1000 nucleotides per second. This machinery is able to build the strand only in one direction from 5' to 3' end. Additionally, synthesis of DNA has to start with a primer - a short fragment of another kind of nucleic acid - RNA. The topology of the double helix replication is shown in Fig. \ref{fig2}. Double helix has two different ends of strands at each end.  Replication starts at the nonrandom point called ,,Origin of replication'' (Ori) and, at the beginning it proceeds only on one old strand starting from its 3' end. Since the synthesized strand is antiparallel - the new strand has 5' end at the start point (it is called ,,leading strand'').\index{leading strand} The second old strand stay free - ,,single stranded'' until the replication move about 2000 nucleotides. Then a starter is synthesized in the replication fork and the second new strand starts to synthesize - it is called ,,lagging strand''. \index{lagging strand}
Meantime the replication of the leading strand proceeds for further 1000 - 2000 nucleotides and another starter for the second fragment of lagging strand is synthesized. Next, starters between DNA fragments of lagging strands are cut off and gaps are filled up with deoxyribonucleotides. There are in fact two different mechanisms of DNA replication - one for the leading strand and the other one for the lagging strand. Of course, during the process of replication the new nucleotides are built into the new strand according to the rules of complementarity: A-T and G-C. There are many reasons in the whole process of replication for which more or less random errors are introduced during the replication when the rules of complementarity are not fulfilled. These errors - called \index{mutation}mutations - are rather rare events, they happen with a frequency of the order of one per genome replication, independently of the genome size. The most frequent errors are substitutions, when the wrong nucleotide is placed in the new strand. For example instead of A opposite to T the G, C or T can be placed. There are three wrong choices for each of four nucleotides. Thus, there are 12 different substitutions. Since the mechanisms of replication of the leading and lagging strands are different, the distributions of the most often substitutions are also different. After many generations, the specific bias of substitutions results in the corresponding bias in the nucleotide composition of the two DNA strands. They are complementary but their nucleotide compositions are different. We call this feature of DNA - ,,DNA asymmetry''. 
It is easy to prove that DNA is asymmetric. You remember Chargaff's rules resulting from the complementarity, called sometimes parity rules \index{parity rule}I: the number of A equals T and the number of G equals C. These rules are deterministic. Let's perform virtual experiment; take a natural DNA sequence, disrupt it for single nucleotides and remake the molecule drawing randomly nucleotides from the whole pool where initially the number A=T and G=C. If we try to build a DNA molecule having such a balanced numbers of nucleotides, using them randomly but keeping the complementarity rules, we can expect that the number of A would equal T and the number of G would equal C in each strand, the differences should be statistically negligible. This stochastic rule is called parity rule \index{parity rule}II. The natural DNA molecules usually do not fulfill the parity rule II - they are asymmetric.
The simplest way to visualize the DNA asymmetry is to perform the DNA walk (Fig. \ref{fig3}). 

\begin{figure}
\centering
\includegraphics[width=0.9\textwidth]{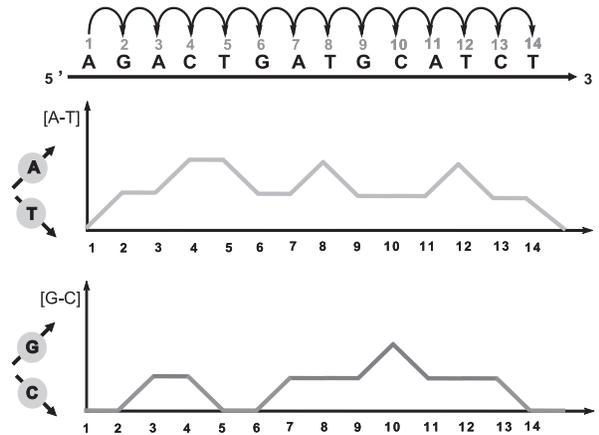}
\caption{\label{fig3} A-T and G-C DNA walks. In the A-T type of a walk, walker starts from the first nucleotide of the analyzed sequence and its moves are: A [1,1], T[1,-1], G[1,0], C[1,0]; in the 
G-C type of walk its moves are: A [1,0], T[1,0], G[1,1], C[1,-1]. DNA walks of this type show local and global asymmetry of DNA molecule.
}
\end{figure}

\begin{figure}
\centering
\includegraphics[width=0.9\textwidth]{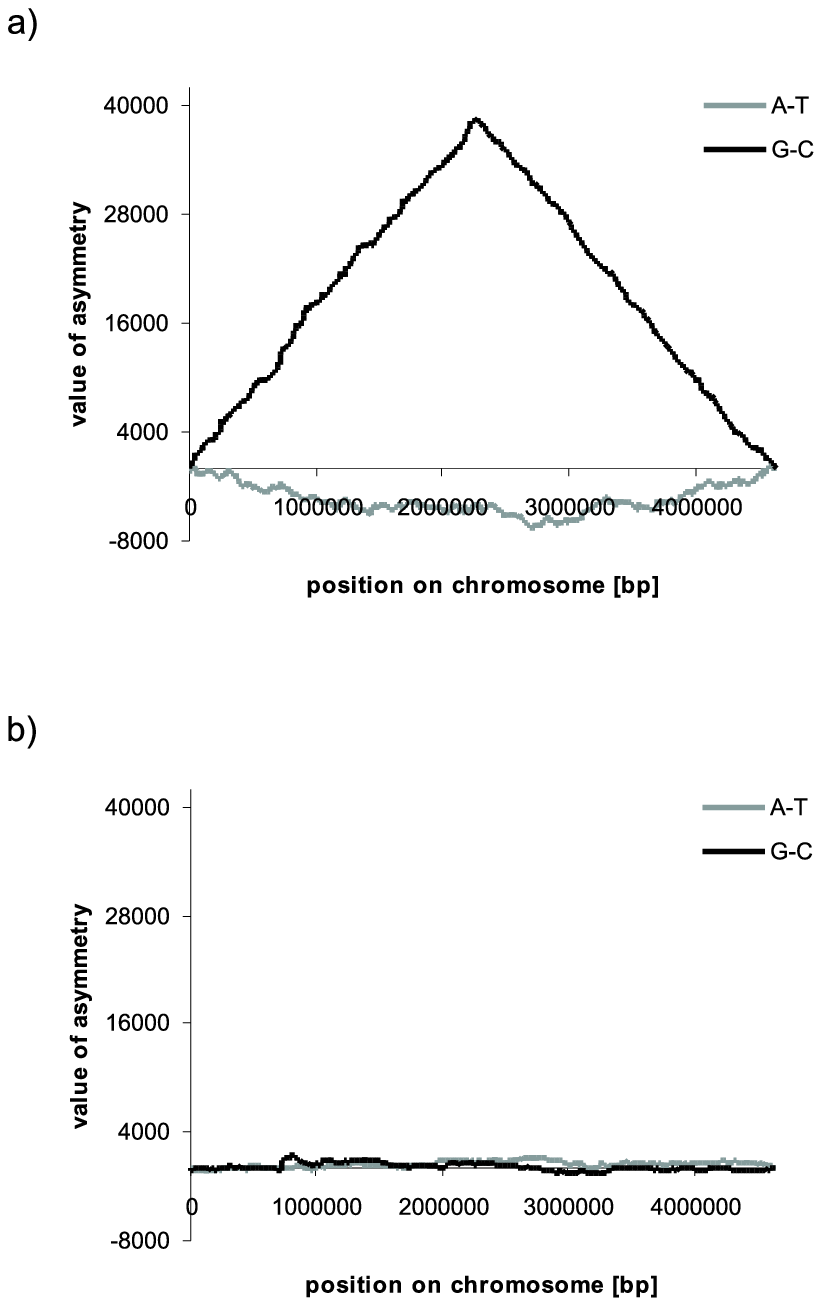}
\caption{\label{fig4} Asymmetry of Escherichia coli DNA chromosome; (a) - A-T and G-C DNA walks performed for the real E. coli chromosome about 4.5 million base pairs long. Walkers start at the terminus of replication, the extremes (very well seen maximum for G-C walk) are in the origin of replication region; (b) - DNA walks performed for the random DNA sequence of the same general nucleotide composition as E. coli chromosome. No significant DNA asymmetry is seen in the same scale. }
\end{figure}

We can put a walker on the first position of the DNA strand and declare that it is moving in the two-dimensional space according to the rules: A [1,1], T[1,-1], G[1,0], C[1,0]. In such a walk, called A/T walk, the walker is going up if there are more A than T in the strand. For the G/C walk the corresponding moves of the DNA walker are: A [1,0], T[1,0], G[1,1], C[1,-1] and walker is going up if there are more G than C in the DNA strand. 
Two examples of the DNA analyses using the DNA walks are presented in Fig. \ref{fig4}. The walk shown in panel b was done for random DNA sequence of the global nucleotide composition characteristic for real E. coli chromosome but randomly distributed along the two strands obeying only the parity rule I \index{parity rule}(complementarity). This is a sequence obtained in our ,,virtual experiment'' described above. The second walk was done for the real E. coli chromosome. The ,,real E. coli chromosome'' means a sequence of nucleotides obtained during the experimental sequencing of E. coli chromosome, (available in the genomic data bases). Note that a sequence of nucleotides of one strand is different than the sequence of the other one but the first strand determines the sequence of the other one unambiguously. Thus, it is enough to put the sequence of only one strand into the data bases. This strand is named the Watson strand, the other one (complementary) is called the Crick strand. The beginning of the circular chromosome is chosen arbitrarily and it is presented in the direction $5'\Rightarrow3'$. It is just for better communication when describing some additional features or sequences on the chromosomes. Comparing the last two DNA walks (randomly arranged sequences and the real one) it is clear that that E. coli chromosome does not obey the parity rule \index{parity rule}II. It is asymmetric \cite{Lobry}. It is divided by Ori and Ter regions for two parts, called replichores. At the points Ori and Ter the asymmetry changes its sign. Again, comparing the topology of replication with the results of DNA walks we can conclude that at the Ori the DNA strands change their role  in replication from the leading to the lagging and vice versa. Concluding; two strands of one double helix DNA molecule are under different mutation pressure and have different nucleotide composition. That is why it is important how genes - fragments of DNA double helix - are positioned on such a molecule and how the information encoded into the DNA is deciphered.

\subsection{Transcription}
\begin{figure}
\centering
\includegraphics[width=0.9\textwidth]{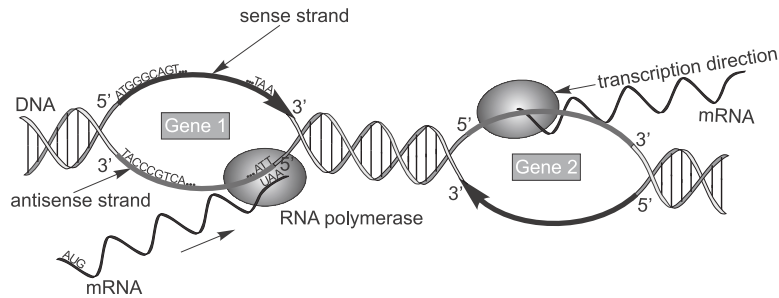}
\caption{\label{fig5} Transcription process. RNA polymerase transcribes the antisense DNA strand. The transcript - RNA sequence - corresponds to the sense DNA strand. Notice that the direction of RNA strand (5'- 3') is the same as the direction of the DNA sense strand. If in the left transcription bubble the sense DNA strand is a leading strand - it is said that the genes is located on the leading strand but notice that the gene transcribed in the left bubble is located on the lagging strand and the direction of its transcription would be different that the direction of replication fork movement. This topology is very important for many mechanisms of gene expression.}
\end{figure}

\index{transcription}
The DNA molecule is a data base of genetic information. To use it, the proper information from this data base should be retrieved, copied and translated into the function. Usually, ,,the function'' is performed by the protein molecule composed of many subunits (up to many thousands) - amino acids. There are 20 different amino acids built into the proteins. We omit the problem of how ,,the proper'' information is found on the huge DNA molecule and we will start just from the point of its copying. To express the desired information encoded in the sequence of deoxyribonucleotides, the very sequence is copied in the process called transcription into another nucleic acid - ribonucleic acid (RNA). There are some differences between DNA and RNA; RNA is single stranded, possesses ribose instead of deoxyribose and Uracil (U) instead of T, complementing with A. RNA is also synthesized from 5' to 3' end on the antiparallel DNA strand (see Fig. \ref{fig5}). 
RNA sequence suppose to have ,,sense'' thus, the antiparallel DNA strand used as matrix for transcription is called antisense strand while the other one (complementary) is again called a sense strand. The position of sense strand of protein coding sequence and its direction describe the topological parameters of gene on chromosome. If the sense strand of gene is located on the leading strand, its transcription has the same direction as the replication fork movement and it is said that ,,the gene'' is located on the leading strand. If the sense strand is located on the lagging strand then the direction of transcription is opposite to the replication fork movement. The RNA sequence is translated into amino acid sequence of proteins. The basis of the complicated decoding system is the genetic code.

\subsection{Genetic code, degeneracy, redundancy}
\begin{figure}
\includegraphics[width=0.8\textwidth]{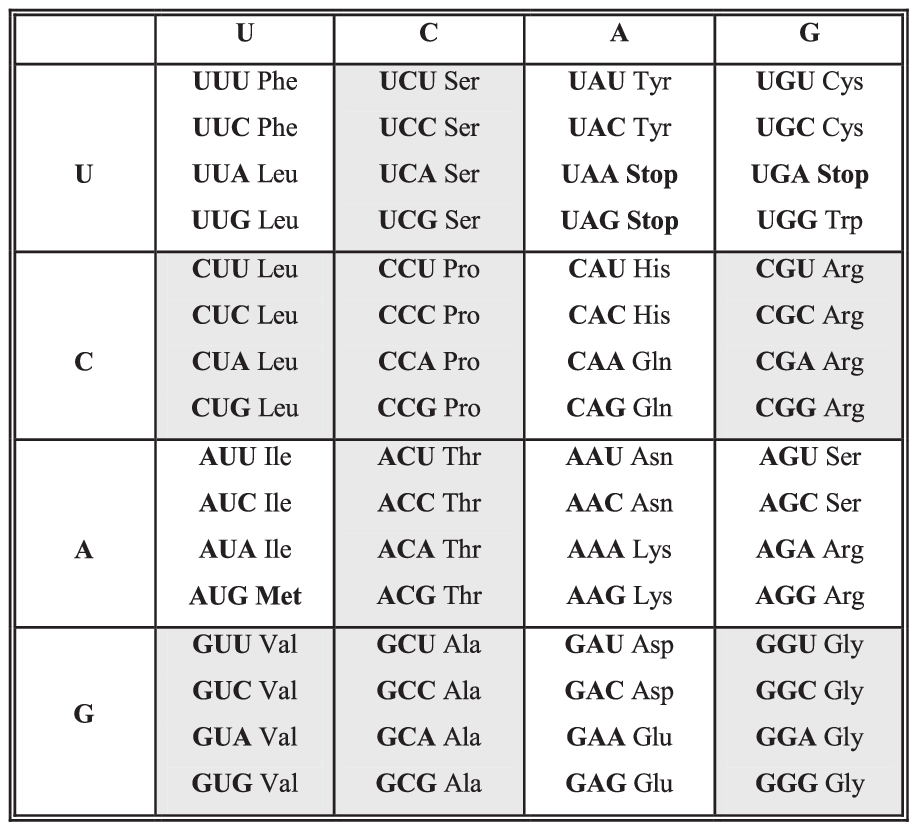}
\caption{\label{tab1} \textbf{Genetic code. }
Notice the way of construction the genetic code table; Codons are grouped in boxes - four codons in each box. Four boxes in one column have the same second nucleotide in codons. Four boxes in one raw have the same first nucleotide in codons. Thus, four codons in one box differ only in the third codon positions. All four codons in each of grey boxes code for the same amino acid (three letters abbreviations of the amino acid names are at the right side of codons). In all other cases pyrimidines in one box (C or U) code for the same amino acid and in only two cases different purines (A or G) change the meaning of codons.
}
\end{figure}

\index{genetic code}\index{genetic code degeneracy}
Nucleic acid is composed of four different subunits while proteins are built of 20 different amino acids. That is why the genetic code has to use at least three nucleotides for coding one amino acid (and really it uses tri-nucleotides) resulting in 64 possible different triplets - much more than the minimum number. The tri-nucleotide coding one amino acid is called ,,codon''\index{codon} (or triplet). The meaning of all codons is shown in the table of genetic code (Fig. \ref{tab1}). 
There are many peculiar properties of the genetic code. First of all the genetic code is universal which means that all organisms in biosphere use the same code. Only some of them, usually with very small genomes, use some ,,dialects'' differing in the sense of single codons. We can state that all organisms communicate in the same language. 
The other properties of genetic code are: not overlapping, without any comas, with the fixed start and stop of translation, unambiguous and degenerate. For our purposes the last two features of the genetic code are very important. Unambiguous means that any codon codes for only one amino acid while degeneration means that the same amino acid can be coded by more than one codon. One of the hypotheses of genetic code evolution assumes that initially the codon assignments varied and it was the selection pressure which optimized the code in respect to reduction of the harmful effects of mutations occurring during replication and transcription and to minimization the frequency of errors during translation process \cite{Sonneborn65}, \cite{Woese} for review see \cite{Freeland03}. During further evolution connected with the increase of genomes, the genetic code was frozen \cite{Crick68} and it was not possible to re-interpret the meaning of codons because every change would affect many amino acid positions in proteins and it would have catastrophic consequences for the organisms. As a result of such an evolution, not only the degeneration of the genetic code is important but also the way it is degenerated. 
One mechanism of optimization results directly from the simple structural relations between nucleotides in the double helix - one large and one small nucleotide fit better to form a pair (a rung in a ladder) and it is much easier to overlook the mutation changing a small nucleotide for another small $(T\leftrightarrow C)$ or large one for another large $(A\leftrightarrow G)$. These mutations\index{mutation} are called transitions. The other kind of mutations, when small nucleotides are replaced by large ones or vice versa (transversions), is much rarer. Note that transitions in the third positions of codons change the sense of codons only in two cases: met/ile and trp/opal. All other transitions in the third positions are synonymous and even half of transversions in the third positions are synonymous. Generally, mutations in the third codon positions are usually accepted by selection. Furthermore, mutations in the first position could be also accepted because they change one amino acid for another one with similar chemical properties (i.e. polarity, hydrophobicity). The most deleterious are mutations in the second positions of codons. Such observations suggest that positions in codons vary in both, their role in coding as well as in the evolution rate, resulting in their nucleotide composition. 

\subsection{Topology of coding sequences}
\begin{figure}
\centering
\includegraphics[width=0.9\textwidth]{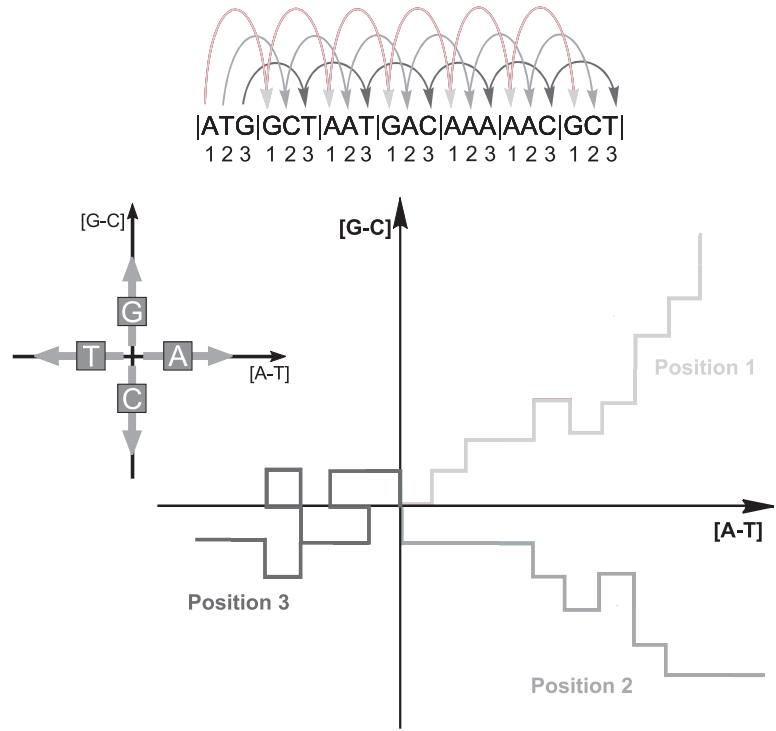}
\caption{\label{fig6} DNA walks on protein coding sequences - spiders. There are three walks in the two-dimensional space. The first walker starts from the first nucleotide, the second walker from the second nucleotide and the third walker starts from the third nucleotide of the first codon. Their jumps every three nucleotides on sequence correspond to moves accordingly to the nucleotide just visited:  G - [0,1], A - [1,0], C - [0,-1] and T - [-1,0]. Each walker produces distinct walk called a spider leg numbered according to the position of nucleotides in analyzed codons.}
\end{figure}

\begin{figure}
\centering
\includegraphics[width=0.85\textwidth]{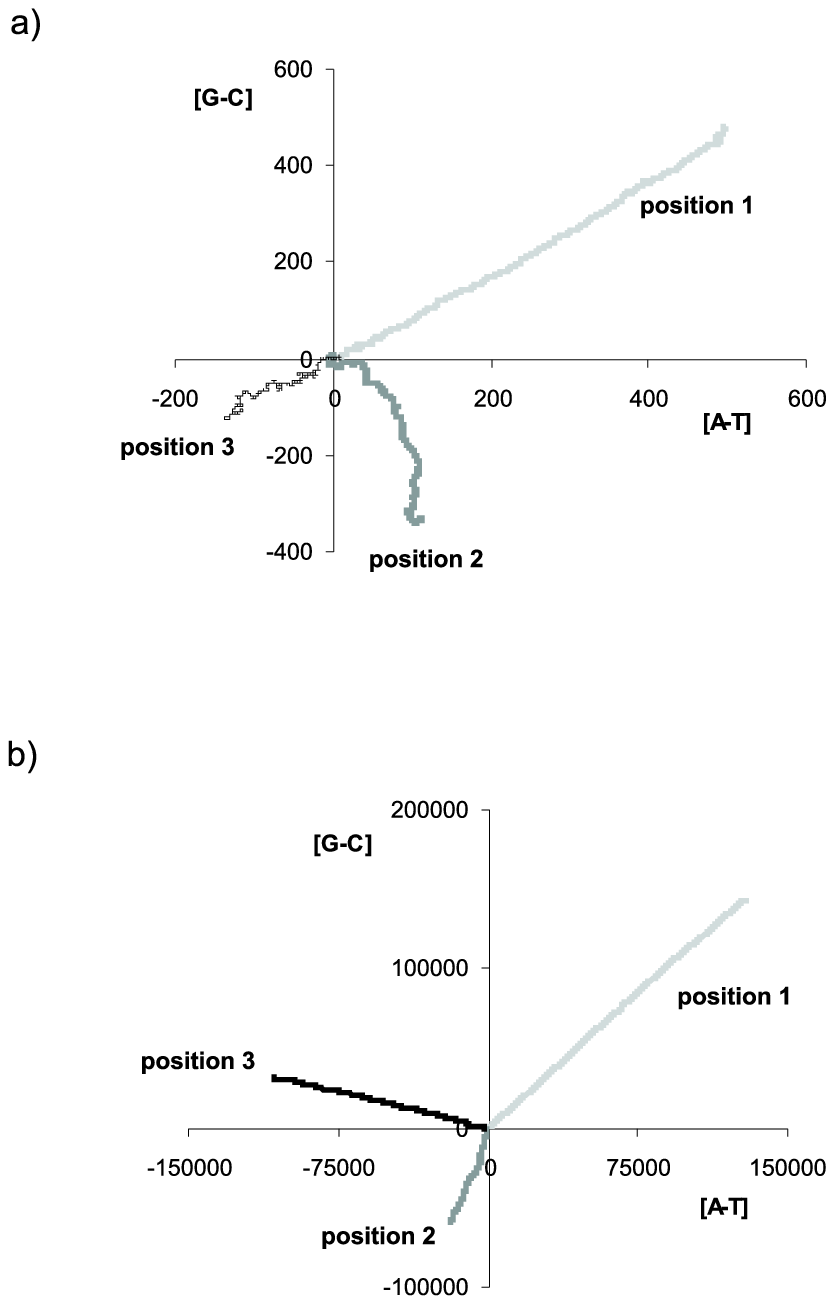}
\caption{\label{fig7} Spiders for a single gene (a) and for all protein coding genes of E. coli located on the leading DNA strand spliced together (b). Straight lines in the (b) panel suggest that the trends in nucleotide composition of all three positions in codons are universal at least for the genes located on this DNA strand.}
\end{figure}

To analyze the nucleotide composition of coding sequences, the DNA walks can be used. In this version of two-dimensional walks \cite{Berthelsen}, the walkers moves are for: G - [0,1], A - [1,0], C - [0,-1] and T - [-1,0]. The best results are obtained when the walks are generated separately for each position in codons (Fig. \ref{fig6}). 
To construct the DNA walk for the first codon positions the walker starts from the first position of the first codon and jumps every three nucleotides until it reaches the first position of the stop codon.\index{stop codon} For the second codon position the walker starts from the second position of the first codon and for the third positions it starts from the third position of the first codon. Three plots describing a coding sequence are called a spider \cite{dudek}. The first example of spider shown in Fig. \ref{fig7} represents three DNA walks done for a single gene of E. coli. It is clear that the walks are different and correspond to different nucleotide composition of the three positions in codons. 
The second spider in Fig. \ref{fig7} shows walks constructed for all genes located on the leading strand of the E. coli genome spliced into one sequence. Now, it is seen that trends for one gene are universal for the whole genome though, in larger scale. More about the analysis of coding capacities of genomes can be found at the web site: \verb'http://www.smorfland.uni.wroc.pl/www/dnawalk.html'.
Such kinds of walks can be done also for analysis of distribution of specific codons \cite{codons}, codons for specific amino acids or groups of amino acids \cite{aminoacids} or any other specific sequences occurring along the chromosomes, like sequences controlling or involved in the replication processes \cite{dnabox}.

\subsection{Mutational pressure}
\label{mutatpres}
The DNA walks done for coding sequences show asymmetry of the coding sequences - a very specific asymmetry - different for each position in codons. It is obvious that this asymmetry is generated by selection which forces the specific amino acid composition of functional proteins and specific codons usage\index{codon usage} - if one amino acid is coded by more than one codon, some preferences of ,,codon usage'' are observed in the genome. A few sections above we have discussed the problem of DNA asymmetry of replichores generated by different mutation pressures for leading and lagging strands. Now it is time to understand what could be the consequence of specific location of coding sequence on chromosome - sense strand of gene located on the leading strand or on the lagging strand. Imagine one example - leading strand is poor in C because this nucleotide is replaced by T preferentially on this strand. On the other hand, the second position of codons is reach in C and this position is very carefully watched by selection. If we put a coding sequence with sense strand reach in C on the leading strand, then it will be very vulnerable for mutations because the probability of substitution C by T is higher and such a substitution could be deleterious for the function of coded protein. 

\section{Evolution of coding sequences}
\begin{figure}[b!h]
\centering
\includegraphics[width=0.6\textwidth]{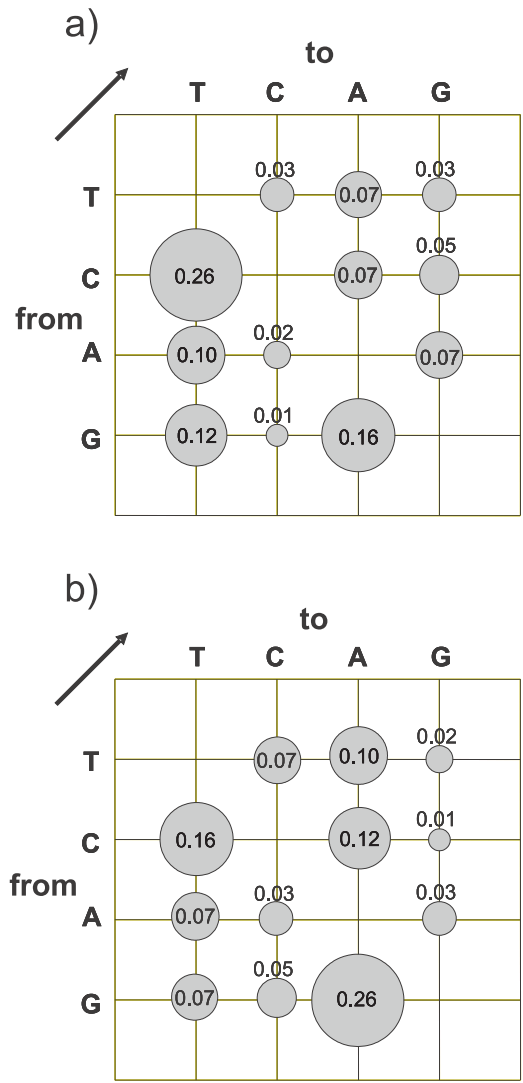}
\caption{\label{tab2} Substitution matrices for the leading (a) and the lagging (b) DNA strands of the Borrelia burgdorferi chromosome. Mutation can replace any nucleotide from the left column by one of the three other nucleotides (upper raw) nevertheless, the substitution probability is different for both, substituted nucleotide (sum of the raw) and substituting nucleotide (numbers in columns). Data in the tables show the relative substitution rate - sum of all numbers equals 1. The real substitution rate in the genome is of the order of million times lower.}
\end{figure}

Coding sequences are under mutation pressure which depends on their location on chromosome and under selection pressure which demands the determined function of protein. Analyzing any genome we see the results of compromise between these two forces. To characterize one force we have to describe the other one. We have succeeded in describing the mutation pressure for the genome of B. burgdorferi \cite{KowalczukTabl}, \cite{Kowalczuk1}, \cite{Kowalczuk2}. In fact there are two, some kind of mirror, matrices describing the relative frequencies of nucleotide substitutions - one for the leading strand and the other one for the lagging strand (Fig. \ref{tab2}). 
The matrices are constructed in such a way that at each line in the first column is a nucleotide which, if chosen for substitution, can be replaced by one of three other nucleotides with a given probability, if it is not substituted it stays unchanged. The sum of all 12 substitution probabilities in the matrix equals one. 
\begin{figure}
\centering
\includegraphics[width=0.9\textwidth]{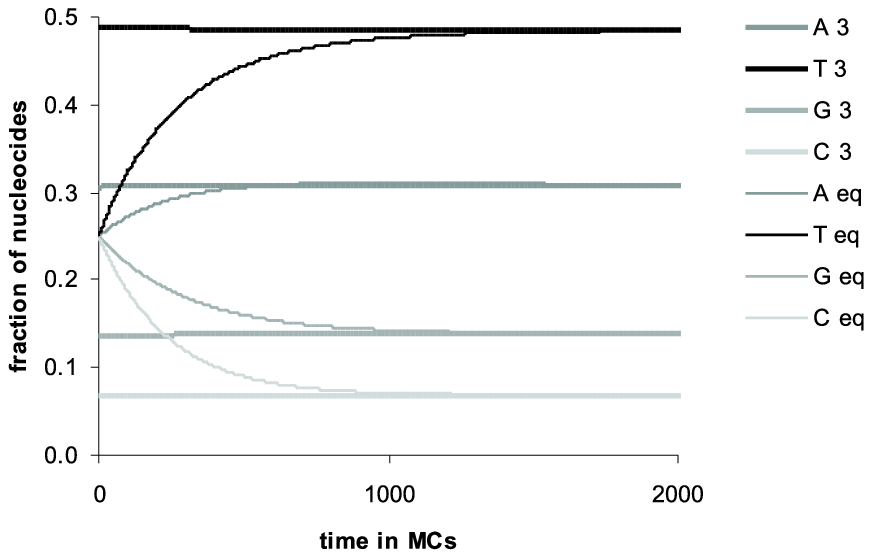}
\caption{\label{fig8} Dynamics of the nucleotide composition changes under mutational pressure characteristic for the leading DNA strand of B. burgdorferi chromosome. There are two different starting compositions of DNA sequences; the first one corresponds to the third position of codons (does not change in the course of simulations) and the second one corresponds to the random sequence with equimolar nucleotide composition (fraction of each nucleotide equals 0.25). The second type of sequence has reached the composition characteristic for the third codon position after about 1000 MCs. }
\end{figure}
Thus, the frequency of substitution in a given sequence depends on the probability of choosing the nucleotide for substitution (parameter $p_{mut}$ - independent of nucleotide composition of the sequence) and on the nucleotide composition of the sequence under the mutation pressure. 
For mutation pressure characteristic for leading strand described by Fig. \ref{tab2}a the highest possible frequency of substitution would be obtained for a sequence composed of C only. If there is no selection pressure, means any constraint set on the sequence under mutation pressure - the sequence would reach the equilibrium composition corresponding to the mutation pressure. 
In Fig. \ref{fig8} we show the dynamics of reaching this state starting from random DNA sequence with the uniform nucleotide composition. The mutational pressure was for leading strand of B. burgdorferi genome. The final nucleotide composition of the DNA sequence corresponds to the nucleotide composition of the third codon position in the coding sequences located on the leading strand of this genome. One can conclude that the third coding positions in this very genome are not under the selection pressure - they are freely shaped by the mutational pressure. The difference (distance) between the nucleotide composition of such sequences, without selection pressure, and any other sequences in the real genome could be considered as a measure of selection force. 

\subsection{Conditions for computer simulation of coding sequences evolution}
\begin{figure}
\centering
\includegraphics[width=0.9\textwidth]{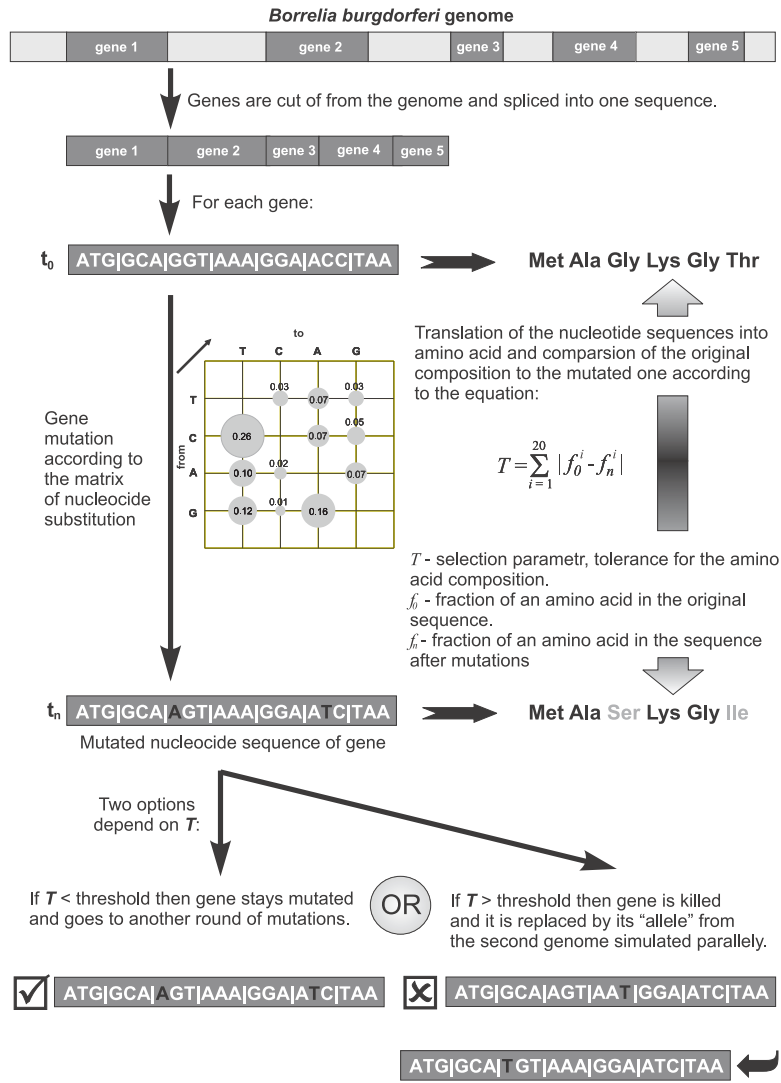}
\caption{\label{fig9} Simulation of the evolution of protein coding sequences. All genes from the B. burgdorferi genomes were cut off the genome and spliced into one sequence. Each gene sequence was translated into amino acid sequence. During each MC step nucleotides were drawn with probability $p_{mut}$ and substituted according to the proper substitution matrix (for leading or lagging strand) and again gene sequences were translated into amino acid sequences. For each of the two amino acid sequences the fractions of amino acid were counted and compared. If the sum of differences was higher than the assumed tolerance - the gene was eliminated and replaced by its homolog from parallel evolving genome otherwise it passed to the next round of simulation. }
\end{figure}

In the previous sections, the definitions and characteristics of coding sequences, genetic code and mutation pressure have been described. The composition of all three positions of codons in the coding sequences usually are different, moreover they differ from those generated by pure mutation pressure because selection pressure eliminates some configurations. To check the effect of mutation pressure on the evolution of coding sequences we have performed computer simulations under following conditions \cite{Kowalczuk3}, \cite{Dudkiewicz}:

\begin{enumerate}
\item Mutational pressure has been described by substitution matrix \index{substitution matrix}for the proper DNA strand, if the coding sequence was located on the leading strand, the used substitution matrix was for the leading strand, if coding sequence was located on lagging strand the substitution matrix was also for lagging strand (\ref{tab2} 2).
\item The original coding sequence was translated into amino acid sequence and fraction of each amino acid has been counted.
\item In one Monte Carlo step, each nucleotide of the gene sequence was drawn with a probability $p_{mut}$, then substituted by another nucleotide with the probability described by the corresponding parameter in the substitution matrix.
\item After each round of mutations, the nucleotide sequences were translated into amino acid sequences and compared to the product of the original gene. Selection pressure was introduced as tolerance for amino acid substitutions. For each gene the selection parameter (T) was calculated. T is the sum of absolute values of differences in fractions of each amino acid between the original sequence ($f_0$) and the sequence after mutations ($f_t$): $T=\sum_{i=1}^{20} | f_0^i - f_t^i |$. It describes deviation in the global amino-acid composition of a protein coded by a given gene after mutations, in comparison to its original sequence from the real genome. If T was below an assumed threshold, the gene stayed mutated and went to the next MC step, if not, the gene was ,,killed'', which means eliminated from the genome and replaced by its allele\index{allele} from the second genomic sequence, originally identical, simulated in the part of the same MC step.
\item The number of all substitutions which occurred during the simulation, the number of accumulated substitutions (accepted), and the number of replacements of genes (killed genes) were counted after each MC step (Fig. \ref{fig9}). 
\end{enumerate}

The maximum tolerance used in these simulations corresponded to the mean value of divergence\index{divergence} between orthologous genes from \textit{B. burgdorferi }and \textit{Treponema pallidum }- a related to \textit{Borrelia}, completely sequenced bacterium \cite{Nowicka}. Orthologus genes mean genes which had a common ancestor sequence in the past. 
Notice that the mutation process in the model is separated from selection by translating the nucleotide sequences into amino acid sequences thus, the important role in the whole evolutionary process is played by the genetic code. 

\subsection{Dynamics of mutation accumulation and gene's elimination}
\begin{figure}
\centering
\includegraphics[width=0.8\textwidth]{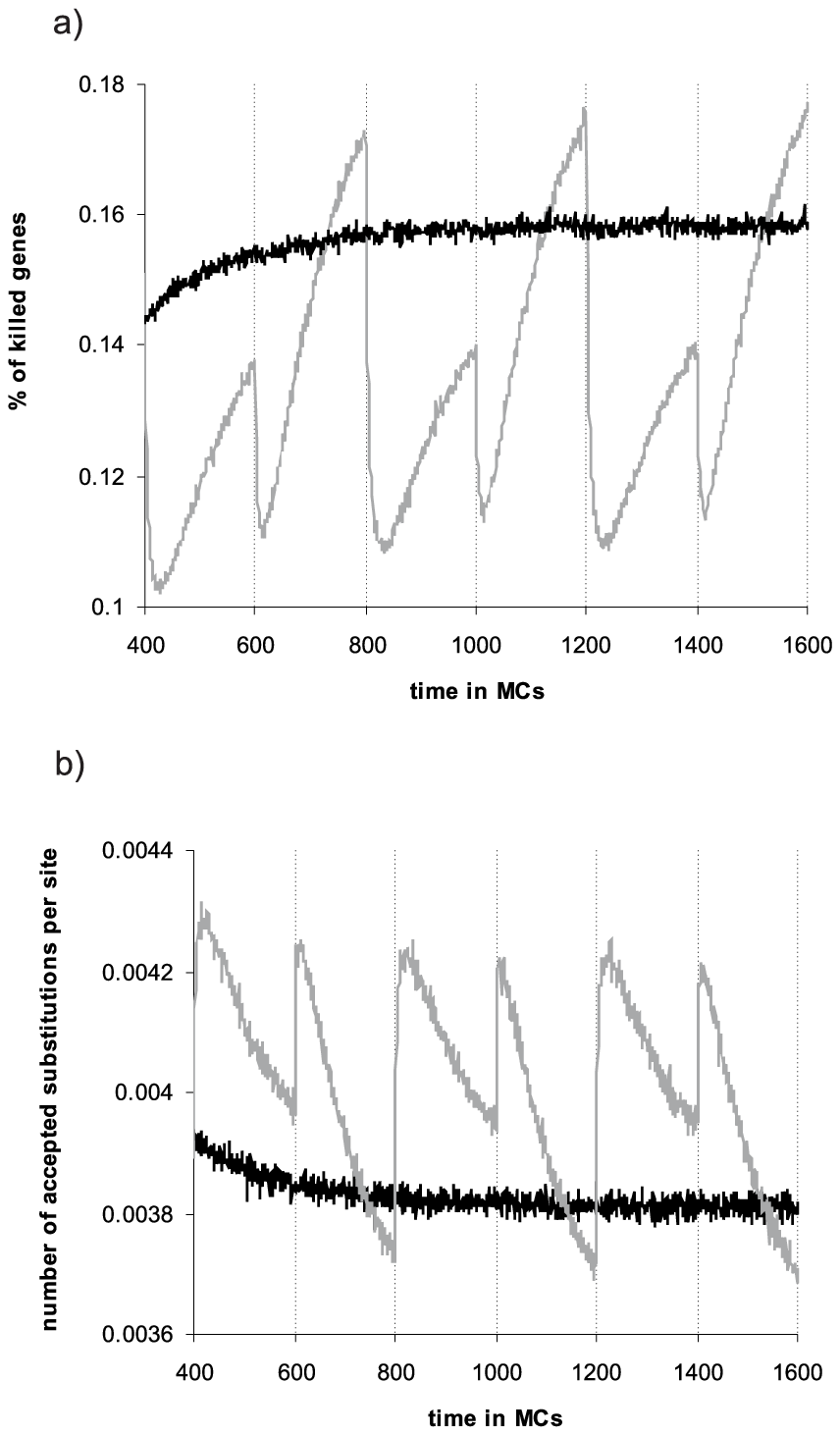}
\caption{\label{fig10} The effect of inversions \index{inversion}on the evolution of coding sequences. Panel (a) the killing effect of mutational pressure; solid, black line without inversions, grey line - with inversions every 200 MCs.
Panel (b) accumulation rate of amino acid substitutions. Notice that each inversion causes drastic decrease of deleterious effect of mutations (killing effect) and simultaneous increase of accepted amino acid substitutions. Generally, inversions decrease the deleterious effect of mutations and increase the evolution rate of coding sequences measured by the amino acid substitution rate (divergence of protein sequences). 
}
\end{figure}

At start of simulations all sequences have the original amino acid composition and T values equals 0. In the course of simulations, mutations accumulate into the nucleotide sequences and, when translated into amino acid sequences, they eventually trespass the assumed tolerance. The nucleotide sequence is then eliminated from the pool and replaced by its orthologous sequence from the parallel evolving sequences. Thus, at the beginning the elimination rate is relatively low and it is growing with time of simulation. This is negatively correlated with the probability of acceptance the mutation into the coding sequence - at the beginning the mutation accumulation rate is relatively high and it is decreasing with time of simulation (Fig. \ref{fig10}). 
This seems to be obvious, but it is some kind of artifact because it has been assumed that at the beginning of simulations all sequences are equally distant from their tolerance limit. Nevertheless, the sequences diverge from the original ones during the evolutions accumulating mutations which are not totally random - mutations are introduced accordingly to the substitution matrix. 
In Nature, it is a known phenomenon of inverting the DNA sequences. If a coding sequence is inverted at the same position (without translocation\index{translocation} to another replichora) it is said that its location is changed from the leading strand to the lagging strand or \textit{vice versa }(see section \ref{mutatpres}). Such inversion results in replacing the mutation pressure characteristic for the leading strand to that of lagging strand (or vice versa) \cite{szczepanik}, \cite{killing}. We have tested the effect of such translocations. When the mutation pressure was changed every MC step, the elimination rate of sequences dropped when compared with stable mutation pressure \cite{Dudkiewicz2}. Moreover, it was associated with the increased accumulation of mutations. The same results were obtained for sequences originally located on the lagging strand. 
Nevertheless, in Nature, the frequency of inversions is rather low and it corresponds to about 200 MCs in simulations under our $p_{mut}$ value \cite{Dudkiewicz3}. We have checked the evolution rate of sequences under such a changing regime of mutation pressure. The results of simulations for sequences originally located on the leading strand are shown in Fig. \ref{fig10}. Notice that inversions\index{inversion} help to escape coding sequences from elimination by selection. The effect is very strong and it is again associated with higher accumulation of mutation in the coding sequences. 
Results of simulations suggest that genomes could use this strategy to lower the costs of mutation pressure and to increase the biodiversity (divergence, mutation accumulation). 
\begin{figure}[t!h]
\centering
\includegraphics[width=0.9\textwidth]{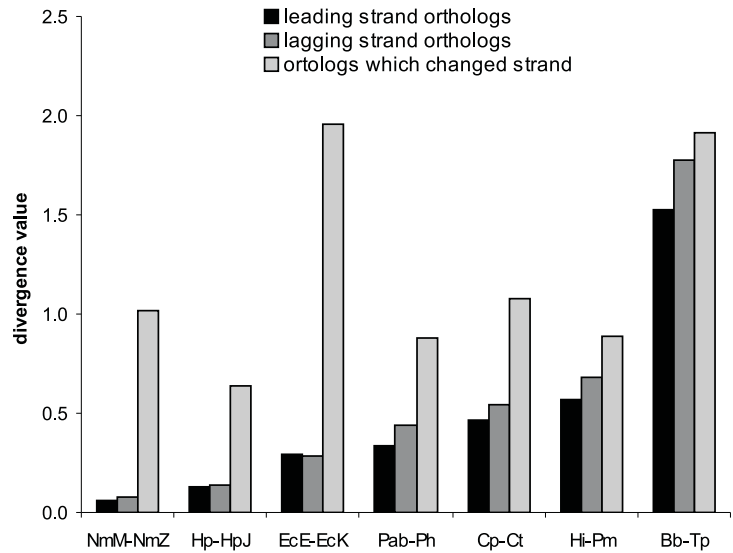}
\caption{\label{fig11} Divergence rate of protein sequences coded by genes located on the leading strands, lagging strands or genes which changed their locations since the speciation.\index{speciation} The first three blocks of columns show the divergence of orthologous sequences in very closely related species or even strains of the same species of bacteria. The difference between the divergence rates of inverted orthologs is much higher than those which haven't change their positions. Abbreviations: \textit{Escherichia coli} K12-MG1655 (EcK) - \textit{E. coli} O157:H7 EDL933 (EcE); \textit{Helicobacter pylori }26695 (Hp) - \textit{H. pylori }J99 (HpJ); 
\textit{Neisseria meningitidis }MC58 (NmM) - \textit{N. meningitidis }Z2491 (NmZ); \textit{Chlamydia pneumoniae} (Cp) - \textit{C. trachomatis} (Ct); \textit{Pyrococcus abyssi }(Pab) - \textit{P. horikoshii }(Ph); \textit{Borrelia burgdorferi }(Bb) - \textit{Treponema pallidum }(Tp); \textit{Haemophilus influenzae }(Hi) - \textit{Pasteurella multocida }(Pm); 
}
\end{figure}
To look for such an effect in Nature we have analyzed the orthologous sequences of some pairs of closely related genomes. We have compared the divergence of corresponding genes in a pair of genomes grouped into classes: in both genomes orthologs stay on the leading DNA strand or, on the lagging DNA strand or, they changed the location - in one genome a sequence is located on the leading strand while in the other one on the lagging strand \cite{szczepanik2}. At very short distances (closely related bacteria strains), we can assume that only one translocation happened during the time of separated evolution. The results of genomic analyses are shown in Fig. \ref{fig11}. 
They are in agreement with the results of simulations. Sequences which were inverted since the divergence of species (or strains) have the highest relative divergence - they were not eliminated and they accumulated more mutations. Sequences which have not been inverted accumulated fewer mutations because they were eliminated by selection more frequently. At larger evolutionary distances this effect is weaker because in each class of sequences we can expect genes which were inverted several times. 

\subsection{The relation between the mutation rate and sequence divergence}
\begin{figure}
\centering
\includegraphics[width=0.8\textwidth]{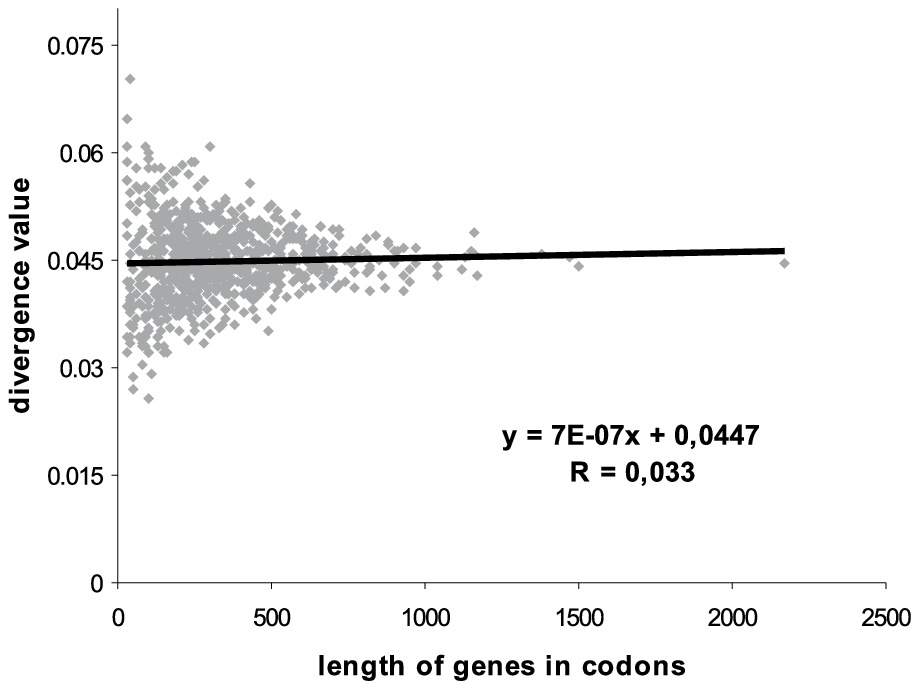}
\caption{\label{fig12} The relation between the divergence rate of coding sequences and their lengths. In Nature there are no significant difference between the length of coding sequences and their divergence rate. To get such an effect in simulations, mutational pressure was lower than 1 mutation per coding sequence in 1 MCs, negative selection for generation the stop codons inside the sequence and elimination of start codons was established, tolerance for amino acid composition was 0.3. }

\vspace{1cm}

\includegraphics[width=0.8\textwidth]{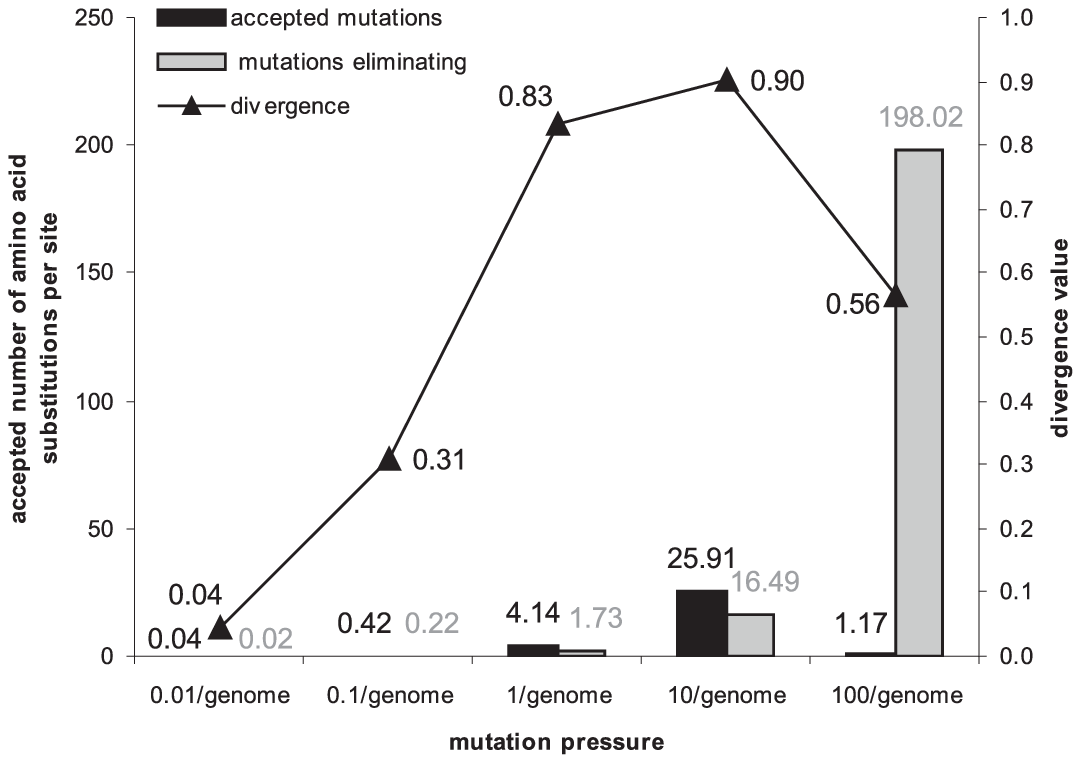}
\caption{\label{fig13} The relations between the mutational pressure and different parameters describing the evolution rate of coding sequences. Notice that the left y-axis shows the total number of accepted (or eliminating) mutations which happened during the simulations. The left scale of y-axis shows the fraction of positions which are different in the coding sequences before and after the simulations (divergence). }
\end{figure}

\index{divergence}
It seems that sequence divergence should be highly correlated with the mutation rate. In fact it is not true. It depends on the tolerance of the sequence. If the limit of tolerance is very small, even single mutation can eliminate the sequence and the mutation accumulation will not be observed. On the other hand, in our simulation we have assumed that all mutations which do not change the meaning of codon are accepted, they are neutral. Thus, it seems that at least all these synonymous mutations\index{synonymous mutations} should be accepted, but it is not true, neither. If mutation rate is too high it could happen that in one sequence two or more substitutions occurred, some of them could be synonymous but other are not tolerated and eliminate the sequence together with the ,,acceptable'' substitutions. There is another effect in the mutation accumulation; if selection tolerates substitution of a given fraction of amino acids in the protein sequence, then the shorter sequences should be more vulnerable to mutations because in long sequences there is still some possibility that mutation in one place could be suppressed by the mutation which occurred at the other place in the ,,inverse direction''. Genomic analyses of the divergence rate of orthologous sequences have shown that the mutation accumulation rate does not depend on the coding sequence length.
Analyzing the vulnerability of coding sequences on nucleotide substitutions it is necessary to consider the generation of stop codons inside the protein coding genes and elimination of start codons. Both kinds of mutations shorten the coding sequences and it is legitimate to assume that they are deleterious and eliminate the gene. The probability of elimination of start codons is independent of the length of coding sequences because each coding sequence possesses only one start codon, while generation of stop codons inside coding sequences linearly depends on the number of codons. These specific properties and effects of substitutions allow estimating some parameters of mutational pressure. Longer sequences are more tolerant for amino acid substitution than shorter ones. On the other hand, the shorter sequences are less vulnerable to the stop codon generation inside them. Thus, it should be possible to find such a set of parameters of mutation rate and tolerance for amino acid substitution where the divergence rate of coding sequences during simulation would not depend on the gene length - just like in natural genomes. In general, the mutation rate should be lower than 1 per selected sequence and tolerance roughly corresponds to that one we have counted on the basis of divergence between B. burgdorferi and T. pallidum - closely related species of bacteria (Fig. \ref{fig12}).

\section{Evolution of whole genomes}
In the previous section, the evolution of single coding sequences was described. Now we would like to describe simulations of evolution of the whole genomes. In particular we would like to answer the questions how the divergence rate depends on the mutation rate and the number of co-evolving genomes. 
We have used the same method of simulations like in case of single coding sequences evolution. The only difference is in the effects of deleterious mutations. In the new approach, mutation which eliminates a single coding sequence eliminates the whole genome. 
The results of simulations showing the effect of different mutational pressure on the divergence rate, accumulation of mutation and elimination of genomes are shown in Fig. \ref{fig13}.

\begin{figure}
\centering
\includegraphics[width=0.9\textwidth]{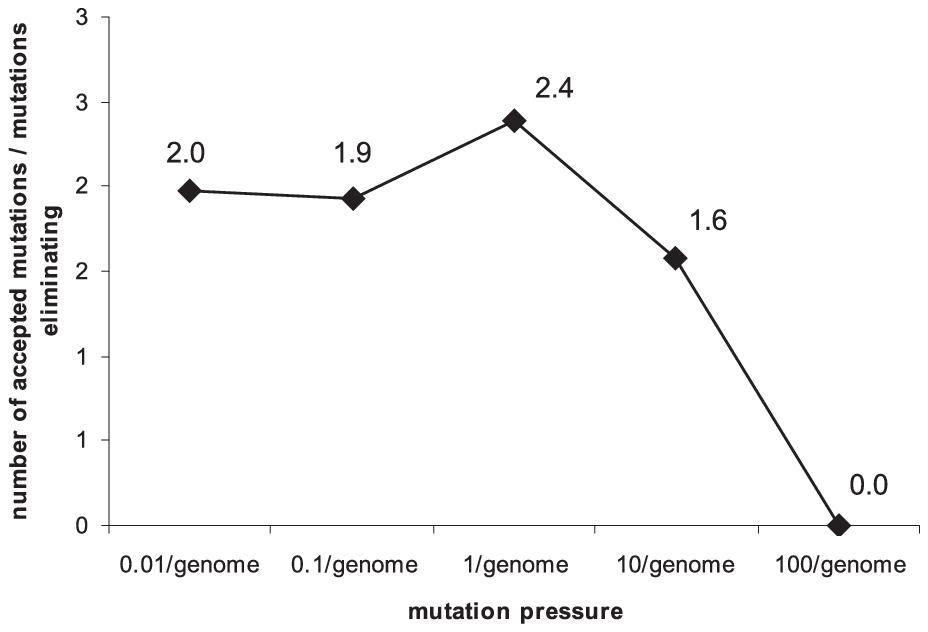}
\caption{\label{fig14} The effect of mutational pressure on the ratio between the number of accepted mutations and eliminating mutations. }

\vspace{1cm}

\centering
\includegraphics[width=0.9\textwidth]{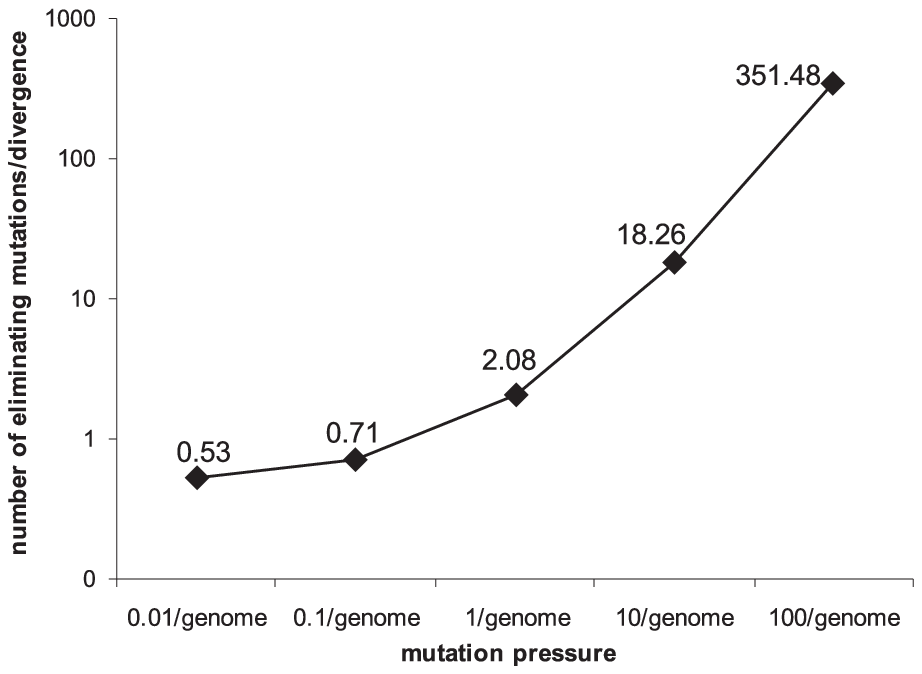}
\caption{\label{fig15} The effect of mutational pressure on the evolutionary costs. It has been assumed that the divergence is a measure of evolution rate and evolutionary costs could be estimated by the ratio between the number of eliminating mutations and divergence. Compare with the results presented in Fig. \ref{fig14}.}
\end{figure}

\begin{figure}
\centering
\includegraphics[width=0.8\textwidth]{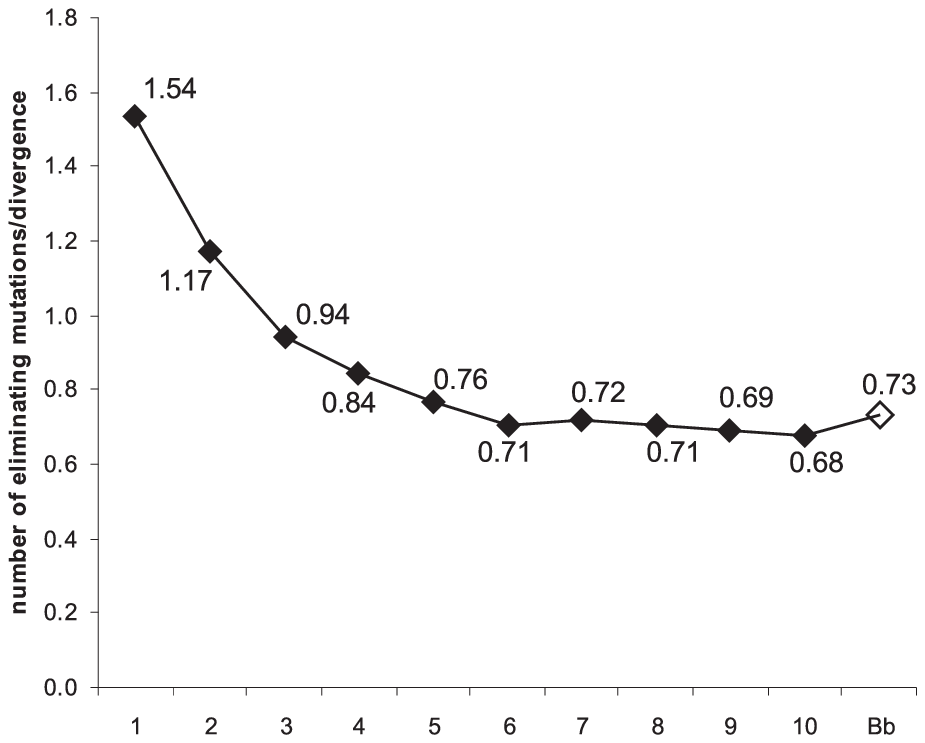}
\caption{\label{fig16} The relation between the evolutionary costs and the genome size. The mutational pressure was constant (1 mutation per genome per MCs).  The smallest ,,genome'' (the first from the left - 1) was represented by integer number (85) of B. burgdorferi genes and it was approximately 10 times smaller than the whole B. burgdorferi genome (the last one at the right). All other ,,genomes'' were multimers of the smallest one.}

\vspace{0.5cm}
\centering
\includegraphics[width=0.8\textwidth]{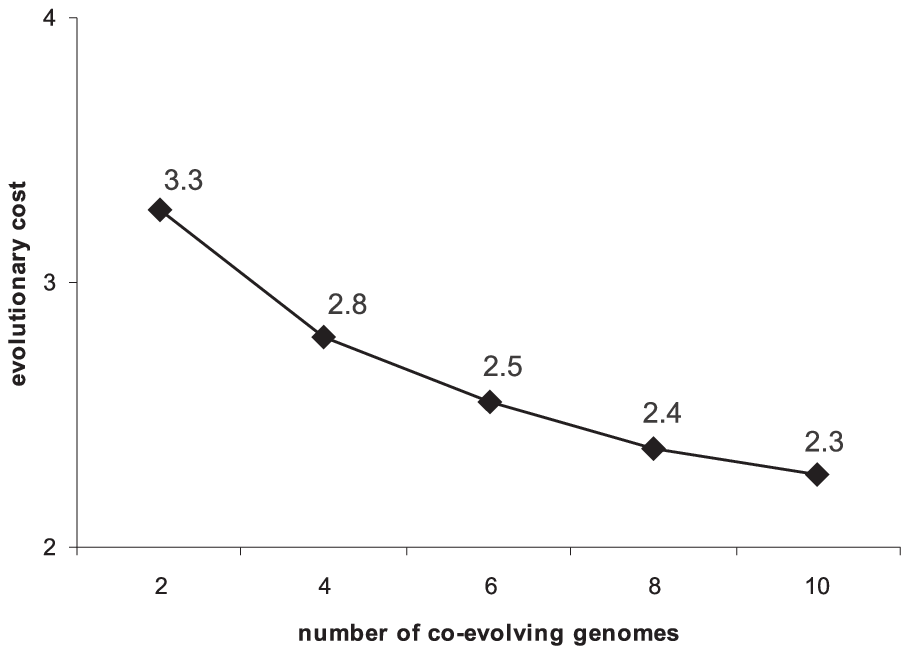}
\caption{\label{fig17} Evolutionary costs for different numbers of co-evolving genomes. Simulations were done for 1 mutation per genome per MCs. Evolutionary costs were estimated by the ratio between the number of eliminating mutations and divergence.}
\end{figure}

Divergence is counted by alignments the original amino acid sequence with a sequence obtained after mutations; it is the fraction of amino acid positions occupied by different amino acids in the compared protein sequences. Notice that even if two different sequences are aligned some positions can be randomly occupied by the same amino acid thus, divergence cannot be higher than 1.0. Results shown in Fig. \ref{fig13} represent the number of accepted amino acid substitutions which have occurred during the whole simulation - many multiple substitutions have been observed; even reversions \index{reversion}were possible - amino acids which have been substituted by other ones have been reintroduced by further mutations at the same position. The number of accepted mutations grows almost linearly with mutational pressure up to 1 mutations in 1 MCs per genome, the acceptance slow down for mutational pressure of the order of 10 per genome per MCs and dramatically decrease with higher mutational pressure.
If the mutational pressure was higher, the number of accidents of elimination the genome by deleterious mutations was also higher and the number of accepted mutations decreased, even if these mutations are neutral. If we accept, as a measure of the evolution costs, the ratio between the number of accepted mutation and the number of mutation eliminating the genome, the optimum of mutational pressure was 1 per genome per generation (Fig. \ref{fig14}, Fig. \ref{fig15}). 
This value is in agreement with some physicists' predictions based on the stability of information as well as with experimental data estimating the mutational rate in different genomes \cite{Azbel}, \cite{Paulo} \cite{Drake}. 
The effectiveness of evolution counted as ratio between the divergence rate and the number of genomes eliminated by deleterious mutations depends also on the genome size. Next series of simulations where performed with the genomes of different sizes - they increased from 0.1 to 1.0 of the original B. burgdorferi genome. The mutational pressure was the same per genome per generation. Results are presented in Fig. \ref{fig16}. 
The evolutionary costs dropped with the genome size approximately twice. If the mutational rate was constant per nucleotide per generation, the evolutionary costs grow with the genome size (not shown). In conclusion, the increase of the genome's size has to be accompanied with the higher accuracy of genetic material replication.
There is one more problem, very important from the evolutionary point of view - how evolution rate depends on the population size. Unfortunately we were not able to simulate the large populations of genomes because such simulations need a lot of computer power. 
Nevertheless, the preliminary studies done with 2 - 10 genomes show that the evolutionary costs decrease with the increase of the population size even if the are no recombination\index{recombination} between the evolving genomes (Fig. \ref{fig17}). It is not in agreement with some outcomes of the Kimura neutral theory \cite{Kimura}. The neutral theory of Kimura could be now revisited using the extensive computer simulations of the whole genomes evolution.

\bibliographystyle{ws-rv-van}

\section*{Acknowledgements}
Calculations have been carried out in Wroc\l{}aw Centre for Networking and Supercomputing (http://www.wcss.wroc.pl), grant \#102.

\printindex                         
\end{document}